\documentclass[aps,tightenlines,nobalancelastpage,floatfix,onecolumn]{revtex4}

\usepackage{graphicx}
\usepackage{amsmath}
\usepackage{amsfonts}
\usepackage{amssymb}
\usepackage{bbm}
\usepackage{epic}

\begin{document}

\title{Advanced action in classical electrodynamics}

\date{\today}

\author{A. D. Boozer} 

\affiliation{
  Norman Bridge Laboratory of Physics 12-33,
  California Institute of Technology,
  Pasadena, CA 91125
}

\begin{abstract}
  The time evolution of a charged point particle is governed by a
  second-order integro-differential equation that exhibits advanced
  effects, in which the particle responds to an external force before
  the force is applied.
  In this paper we give a simple physical argument that clarifies the
  origin and physical meaning of these advanced effects, and we
  compare ordinary electrodynamics with a toy model of electrodynamics
  in which advanced effects do not occur.
\end{abstract}


\maketitle

\section{Introduction}

Charged point particles in classical electrodynamics obey an equation
of motion, known as the Lorentz-Dirac equation, that is third order in
proper time, and therefore presents various difficulties
\cite{rohrlich, barut, jackson, dirac}.
In particular, the Lorenz-Dirac equation admits unphysical runaway
solutions, in which the particle accelerates without bound even in the
absence of an external driving force.
The runaway solutions can be eliminated by replacing the Lorenz-Dirac
equation with a second-order integro-differential equation, but this
equation has problems of its own: it exhibits advanced effects, in
which the particle responds to an external force before the force has
been applied.

In this paper we give a simple physical argument that clarifies the
origin and physical meaning of these advanced effects, and we compare
ordinary electrodynamics with a toy model of electrodynamics in which
advanced effects do not occur.
The toy model is closely analogous to ordinary electrodynamics; for
example, there are fields that are analogous to the electric and
magnetic fields, and these fields both mediate an interaction between
charged particles and support freely propagating radiation.
The toy model, however, does not exhibit the conceptual problems that
plague ordinary electrodynamics; for example, the
self-energy of a point particle does not diverge, and there are no
advanced effects.
Thus, the toy model shows that it is possible to construct a
mathematically consistent theory of coupled particles and fields
that is not subject to these pathologies.
By performing a detailed comparison between the toy model and
ordinary electrodynamics, we can gain insight into the
conceptual foundations of electrodynamics, and into the meaning of the
advanced effects.

The paper is organized as follows.
In section
\ref{sec:lorentz-dirac},
we briefly review the Lorentz-Dirac equation of motion, the
corresponding integro-differential equation of motion, and the
causality problems associated with it.
In section
\ref{sec:extended-particle}, we consider a spatially extended charged
particle, and give a simple physical argument that explains the origin
of the advanced effects.
In section
\ref{sec:toy-model},
we introduce the toy model of electrodynamics.
In section
\ref{sec:toy-model-advanced},
we consider a spatially extended particle in the toy model, and derive
analogs to the various equations of motion for electrodynamics.

The following notation is used in this paper.
The function $\epsilon(x)$ is the sign function, defined such that
$\epsilon(x) = 1$ if $x > 0$, $\epsilon(x) = 0$ if $x = 0$, and
$\epsilon(x) = -1$ if $x < 0$.
The function $\theta(x)$ is the step function, defined such that
$\theta(x) = 1$ if $x > 0$, $\theta(x) = 1/2$ if $x = 0$, and
$\theta(x) = 0$ if $x < 0$.
The metric tensor $\eta_{\mu\nu}$ is defined such that
$\eta_{00} = 1$, $\eta_{11} = \eta_{22} = \eta_{33} = -1$.
The three-vector component of a four-vector is denoted by boldface
type; for example, $x^\mu = (x^0, \mathbf{x})$.

\section{The Lorentz-Dirac equation}
\label{sec:lorentz-dirac}

Consider a point particle coupled to the electromagnetic field.
We will let $m$ and $e$ denote the mass and charge of the particle,
and we will let $z^\mu(\tau)$ denote the position of the particle at
proper time $\tau$.
Also, we will define $v^\mu = dz^\mu/d\tau$ to be the velocity of
the particle and $a^\mu= dv^\mu/d\tau$ 
to be its acceleration.
The equation of motion for the particle is
\begin{eqnarray}
  \label{eqn:particle-eqn-of-motion}
  m a^\mu = K_f^\mu + K_{ext}^\mu,
\end{eqnarray}
where $K_f^\mu$ is the force exerted on the particle by the
electromagnetic field and $K_{ext}^\mu$ is an arbitrary externally
imposed force.
For simplicity, we will assume that $K_{ext}^\mu$ depends only on the
proper time $\tau$, and not on the position or velocity of the
particle.
The force $K_f^\mu$ is given by
\begin{eqnarray}
  K_f^\mu(\tau) = e F^{\mu\nu}(z(\tau))\,v_\nu(\tau),
\end{eqnarray}
where
$F^{\mu\nu} = \partial^\mu A^\nu - \partial^\nu A^\mu$ is the
electromagnetic field strength tensor and $A^\mu$ is the vector
potential.
In the Lorentz gauge ($\partial_\mu A^\mu = 0$), the vector potential
satisfies the field equation
\begin{eqnarray}
  \label{eqn:field-eqn}
  \Box A^\mu = 4\pi J^\mu,
\end{eqnarray}
where $J^\mu(x)$, the current density, is given by
\begin{eqnarray}
  \label{eqn:current-density-point}
  J^\mu(x) = e\int v^\mu(\tau)\,\delta^{(4)}(x - z(\tau))\,d\tau.
\end{eqnarray}
Equations
(\ref{eqn:particle-eqn-of-motion}--\ref{eqn:current-density-point})
give a complete description of the coupled particle-field system.

We can decompose each solution to the field equation
(\ref{eqn:field-eqn}) into the sum of an inhomogeneous solution, which
describes the potential generated by the particle, and a homogeneous
solution, which describes freely propagating radiation.
It is useful to perform this decomposition in two different ways:
\begin{eqnarray}
  \label{eqn:A-decomposition}
  A^\mu = A_r^\mu + A_{in}^\mu = A_a^\mu + A_{out}^\mu,
\end{eqnarray}
where $A_r^\mu$ and $A_a^\mu$ are the retarded and advanced
potentials generated by the particle, and $A_{in}^\mu$ and
$A_{out}^\mu$ describe incoming and outgoing radiation.
The retarded and advanced potentials are given by
\begin{eqnarray}
  \label{eqn:A-ret-adv}
  A_r^\mu(x) = \int D_r(x - x')\,J^\mu(x')\,d^4 x', \qquad
  A_a^\mu(x) = \int D_a(x - x')\,J^\mu(x')\,d^4 x',
\end{eqnarray}
where $D_r(x)$ and $D_a(x)$ are the retarded and advanced Green
functions for the inhomogeneous wave equation:
\begin{eqnarray}
  D_r(x) =
  2\,\theta(x^0)\,\delta(x \cdot x) =
  |\mathbf{x}|^{-1}\,\delta(x^0 - |\mathbf{x}|), \qquad
  D_a(x) =
  2\,\theta(-x^0)\,\delta(x \cdot x) =
  |\mathbf{x}|^{-1}\,\delta(x^0 + |\mathbf{x}|).
\end{eqnarray}
Using the decompositions of the vector potential given in equation
(\ref{eqn:A-decomposition}), we can express the field-strength
tensor and the electromagnetic force as
\begin{eqnarray}
  F^{\mu\nu} =
  F_r^{\mu\nu} + F_{in}^{\mu\nu} =
  F_a^{\mu\nu} + F_{out}^{\mu\nu}, \qquad
  K_f^\mu = K_r^\mu + K_{in}^\mu = K_a^\mu + K_{out}^\mu,
\end{eqnarray}
where, for example,
\begin{eqnarray}
  \label{eqn:F-ret-K-ret}
  F_r^{\mu\nu} =
  \partial^\mu A_r^\nu - \partial^\nu A_r^\mu, \qquad
  K_r^\mu(\tau) = e F_r^{\mu\nu}(z(\tau))\,v_\nu(\tau).
\end{eqnarray}
Physically, $K_{in}$ describes the force exerted on the particle by
incoming radiation, and $K_r$ describes the self-force exerted on the
particle by its own retarded field.
In what follows, we will assume that there is no incoming radiation,
so $K_{in}^\mu = 0$ and $K_f^\mu = K_r^\mu$.

We can obtain an explicit expression for the self-force by combining
equations (\ref{eqn:current-density-point}), (\ref{eqn:A-ret-adv}),
and (\ref{eqn:F-ret-K-ret}).
To perform the calculation it is useful to express the self-force
in the form $K_r^\mu = K_+^\mu + K_-^\mu$, where
$K_\pm^\mu \equiv (1/2)(K_r^\mu \pm K_a^\mu)$.
As was first shown by Dirac \cite{dirac} (see also \cite{barut},
pages 187--9), the component $K_-^\mu$ is
well-defined and is given by
\begin{eqnarray}
  \label{eqn:K-minus}
  K_-^\mu = m\tau_0(\dot{a}^\mu + (a\cdot a)v^\mu),
\end{eqnarray}
where $\dot{a}^\mu \equiv da^\mu/d\tau$ and
$\tau_0 \equiv (2/3)(e^2/m)$.
The component $K_+^\mu$, however, is infinite.
The reason for this can be traced to the fact that the retarded fields
of a point particle diverge as we approach the particle;
for a spatially extended particle, one can show that $K_+^\mu$ is
finite and is given by
\begin{eqnarray}
  \label{eqn:K-plus}
  K_+^\mu = -m_S a^\mu + \cdots
\end{eqnarray}
where $m_S$ is the self-energy of the particle and the dots indicate
additional terms that vanish in the point particle limit.
We derive an expression for the self-energy of a spherically symmetric
particle in Appendix \ref{sec:appendix-self-energy}, but for now we
simply note that the self-energy scales like $1/\sigma$, where
$\sigma$ is the particle size, and therefore diverges in the point
particle limit.
Thus, in order to obtain a finite expression for the self-force, let
us assume that the particle is spatially extended, but small enough
that $K_-^\mu$ is well-approximated by equation (\ref{eqn:K-minus})
and $K_+^\mu$ is well-approximated by the first term of equation
(\ref{eqn:K-plus}):
\begin{eqnarray}
  \label{eqn:K-r}
  K_r^\mu = 
  K_-^\mu + K_+^\mu = 
  m\tau_0(\dot{a}^\mu + (a\cdot a)v^\mu) - m_S a^\mu.
\end{eqnarray}
Substituting this result into equation
(\ref{eqn:particle-eqn-of-motion}), we find that the equation of
motion for the particle is
\begin{eqnarray}
  m a^\mu =
  m\tau_0(\dot{a}^\mu + (a\cdot a)v^\mu) - m_S a^\mu + K_{ext}^\mu.
\end{eqnarray}
Here $m$ is the bare mass of the particle; that is, the mass that the
particle would have if its charge were set to zero.
Let us define a renormalized mass $m_R \equiv m + m_S$ and a
renormalized time constant
$\tau_R \equiv (m/m_R)\tau_0 = (2/3)(e^2/m_R)$.
We can then express the equation of motion as
\begin{eqnarray}
  \label{eqn:lorentz-dirac}
  a^\mu - \tau_R \dot{a}^\mu =
  \tau_R (a\cdot a)v^\mu + (1/m_R) K_{ext}^\mu.
\end{eqnarray}
This is the Lorentz-Dirac equation.
We derived this equation by considering a small spatially extended
particle, but it is well-defined in the point particle limit, provided
we hold $m_R$ constant and allow $m$ to diverge in order to compensate
for the divergence of $m_S$.

It is instructive to consider the low-velocity limit of the
Lorentz-Dirac equation:
\begin{eqnarray}
  \label{eqn:abraham-lorentz}
  \mathbf{a} - \tau_R\dot{\mathbf{a}} =
  (1/m_R)\mathbf{K}_{ext},
\end{eqnarray}
where $\dot{\mathbf{a}} \equiv d\mathbf{a}/dt$.
An ordinary equation of motion is second order in time, and thus
requires initial conditions $\mathbf{z}_0$ and $\mathbf{v}_0$ for the
position and velocity, but since equation
(\ref{eqn:abraham-lorentz}) is third order in time we need an
additional initial condition $\mathbf{a}_0$ for the acceleration.
If $\mathbf{a}_0$ is not chosen properly we obtain unphysical runaway
solutions.
We can see this for the case of a free particle
($\mathbf{K}_{ext} = 0$), for which the solution to
(\ref{eqn:abraham-lorentz}) is
\begin{eqnarray}
  \mathbf{a}(t) = \mathbf{a}_0\,e^{t/\tau_R}.
\end{eqnarray}
Thus, unless $\mathbf{a}_0 = 0$ we obtain a runaway solution in which
the acceleration increases exponentially in time.
We can eliminate the runaway solutions by expressing the solution to
(\ref{eqn:abraham-lorentz}) in terms of a Green function $G(t)$:
\begin{eqnarray}
  \label{eqn:abraham-lorentz-int-diff}
  \mathbf{a}(t) =
  (1/m_R)\int G(t-t')\,\mathbf{K}_{ext}(t')\,dt',
\end{eqnarray}
where $G(t)$ is defined such that $G(t) \rightarrow 0$ for
$t \rightarrow \pm \infty$ and
\begin{eqnarray}
  \label{eqn:define-G}
  G(t) - \tau_R\,\dot{G}(t) = \delta(t).
\end{eqnarray}
Let $\tilde{G}(\omega)$ denote the Fourier transform of $G(t)$.
From equation (\ref{eqn:define-G}), it follows that
\begin{eqnarray}
  \tilde{G}(\omega) =
  \int G(t)\,e^{i\omega t}\,dt =
  (1 + i\tau_R\omega)^{-1}, \qquad
  G(t) =
  (2\pi)^{-1}\int \tilde{G}(\omega)\,e^{-i\omega t}\,d\omega =
  (1/\tau_R)\theta(-t)\,e^{t/\tau_R}.
\end{eqnarray}
We can view equation (\ref{eqn:abraham-lorentz-int-diff}) as a new
equation of motion for the particle.
By its construction it is a solution to equation
(\ref{eqn:abraham-lorentz}), so integrating equation
(\ref{eqn:abraham-lorentz-int-diff}) subject to the initial conditions
$\{\mathbf{z}_0, \mathbf{v}_0\}$ is equivalent to integrating equation
(\ref{eqn:abraham-lorentz}) subject to the initial conditions
$\{\mathbf{z}_0, \mathbf{v}_0, \mathbf{a}_0\}$, where $\mathbf{a}_0$
is given by
\begin{eqnarray}
  \label{eqn:a0}
  \mathbf{a}_0 =
  (m_R \tau_R)^{-1}
  \int_0^\infty e^{-t'/\tau_R}\,\mathbf{K}_{ext}(t')\,dt'.
\end{eqnarray}
In effect, the integro-differential equation sets the initial
condition $\mathbf{a}_0$ so as to eliminate the runaway solutions.
We can see this for the case of a free particle: when
$\mathbf{K}_{ext} = 0$,  equation (\ref{eqn:a0}) sets
$\mathbf{a}_0 = 0$.
Note that since $G(t) > 0$ for $t < 0$, equation
(\ref{eqn:abraham-lorentz-int-diff}) exhibits advanced effects, in
which the acceleration of the particle at time $t$ depends on the
value of the external force at times $t' > t$.

For simplicity, we have shown how the runaway solutions can be
eliminated in the low-velocity limit, but the same method can be
applied to the full relativistic Lorenz-Dirac equation (see section
6.6 of \cite{rohrlich}, section V.6 of \cite{barut}, and
\cite{rohrlich-1961}).
The resulting equation of motion is
\begin{eqnarray}
  \label{eqn:lorentz-dirac-int-diff}
  a^\mu(\tau) =
  \int_\tau^\infty e^{-(\tau' - \tau)/\tau_R}\,
  ((a(\tau') \cdot a(\tau'))\,v^\mu(\tau') +
  (m_R\tau_R)^{-1} K_{ext}^\mu(\tau'))\,d\tau'.
\end{eqnarray}
Note that $a^\mu(\tau)$ depends on the value of $K_{ext}^\mu(\tau')$
at proper times $\tau' > \tau$, so equation
(\ref{eqn:lorentz-dirac-int-diff}) also exhibits advanced effects.

\section{Extended particles in electrodynamics}
\label{sec:extended-particle}

\begin{figure}[h]
  \centering
  \includegraphics{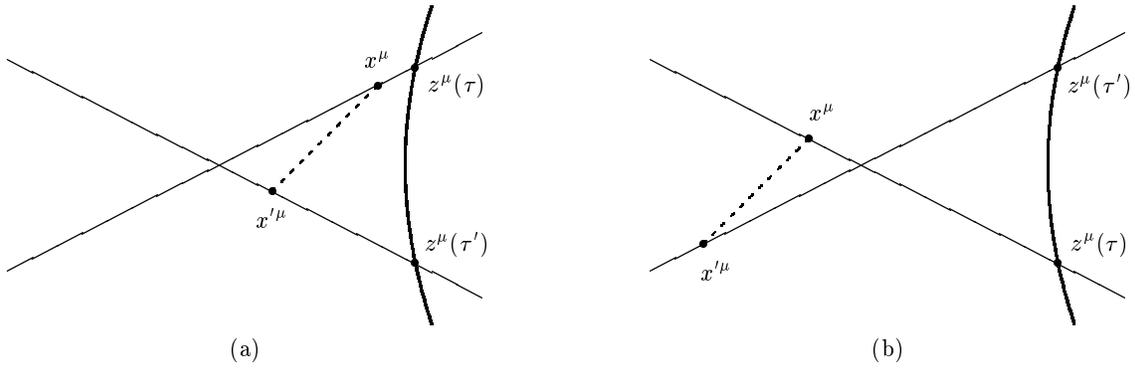}
  \caption{
    \label{fig:spacetime-diagrams-electrodynamics}
    Spacetime diagrams, which illustrate pairs of events $x^\mu$ and
    $x'^\mu$ that give nonzero contributions to
    $\bar{K}_r^\mu(\tau,\tau')$.
    The thick solid lines indicate the particle trajectory,
    the thin solid lines indicate the planes of simultaneity for
    proper times $\tau$ and $\tau'$,
    the dashed lines indicate the light-like interval between events
    $x^\mu$ and $x'^\mu$.
    (a) Conditions (1--3) are satisfied for $\tau > \tau'$,
    corresponding to retarded effects.
    (b) Conditions (1--3) are satisfied for $\tau < \tau'$,
    corresponding to advanced effects.
  }
\end{figure}

We can gain some insight into the origin of the advanced effects
by generalizing equation (\ref{eqn:lorentz-dirac-int-diff})
to the case of a spatially extended particle of arbitrary size.
For simplicity, let us consider an extended particle that is
spherically symmetric.
The current density is then given by
\begin{eqnarray}
  J^\mu(x) =
  e\int (1 - a(\tau) \cdot (x - z(\tau)))\,
  f(-(x - z(\tau))^2)\,v^\mu(\tau)\,
  \delta(v(\tau) \cdot (x - z(\tau)))\,d\tau,
\end{eqnarray}
where $e$ is the total charge of the particle and $f(r^2)$ describes
the radial charge distribution (this expression for the current
density is derived in section 7.4 of \cite{rohrlich}).
It is convenient to express the current density in the form
\begin{eqnarray}
  J^\mu(x) = \int \bar{J}^\mu(x,\tau)\,d\tau, \quad
\end{eqnarray}
where
\begin{eqnarray}
  \label{eqn:j-bar}
  \bar{J}^\mu(x,\tau) \equiv
  e(1 - a(\tau) \cdot (x - z(\tau)))\,
  f(-(x - z(\tau))^2)\,v^\mu(\tau)\,
  \delta(v(\tau) \cdot (x - z(\tau))).
\end{eqnarray}
Note that $\bar{J}^\mu(x,\tau)$ vanishes unless $x^\mu$ lies in the
plane of simultaneity for proper time $\tau$.
We can describe the coupling of the particle to the electromagnetic
field in terms of an action $S_i$ and Lagrangian $L_i$:
\begin{eqnarray}
  S_i = -\int A^\mu(x)\,J_\mu(x)\,d^4 x = \int L_i\,d\tau, \qquad
  L_i = -\int A^\mu(x)\,\bar{J}_\mu(x,\tau)\,d^4 x.
\end{eqnarray}
From the Euler-Lagrange equations, it follows that the electromagnetic
force on the extended particle is
\begin{eqnarray}
  K_f^\mu(\tau) =
  \int F^{\mu\nu}(x)\,\bar{J}_\nu (x,\tau)\,d^4 x.
\end{eqnarray}
As before, we can separate the electromagnetic force into a component
$K_r^\mu$ that describes the self-force and a component $K_{in}^\mu$
that describes the force exerted on the particle by incoming
radiation.
Using equation (\ref{eqn:A-ret-adv}) to solve for $A_r^\mu(x)$ in
terms of the current density, we find that the self-force is given by
\begin{eqnarray}
  \label{eqn:K-ret-extended}
  K_r^\mu(\tau) =
  \int F_r^{\mu\nu}(x)\,\bar{J}_\nu (x,\tau)\,d^4 x =
  \int \bar{K}_r^\mu(\tau,\tau')\,d\tau',
\end{eqnarray}
where we have defined
\begin{eqnarray}
  \label{eqn:K-ret-extended-bar}
  \bar{K}_r^\mu(\tau,\tau') \equiv
  \int\!\!\!\int
  \bar{J}^\alpha (x,\tau)\,{\Delta^\mu}_{\alpha\beta}(x,x')\,
  \bar{J}^\beta (x',\tau')
  \,d^4 x\,d^4 x', \qquad
  {\Delta^\mu}_{\alpha\beta}(x,x') \equiv
  (\eta_{\alpha\beta}\,\partial^\mu - {\eta^\mu}_\beta\,\partial_\alpha)
  D_r(x - x').
\end{eqnarray}
If we assume that there is no incoming radiation,
the equation of motion for the extended particle is
\begin{eqnarray}
  \label{eqn:lorenz-dirac-extended}
  m a^\mu = K_r^\mu + K_{ext}^\mu,
\end{eqnarray}
where $K_r^\mu$ is given by equation (\ref{eqn:K-ret-extended}) and
$K_{ext}^\mu$ describes an arbitrary externally imposed force.
This is a second-order integro-differential equation; it is exact, and
holds for extended particles of arbitrary size.
The equation of motion
(\ref{eqn:lorentz-dirac-int-diff}) that we derived in the
previous section should approximate this equation of motion in the
limit of a small particle
(the parameters $m_R = m + m_S$ and $\tau_R = (2/3)(e^2/m_R)$ that
appear in equation (\ref{eqn:lorentz-dirac-int-diff}) are set by the
value of the self-energy $m_S$, which can be determined from $f(r^2)$
using equation (\ref{eqn:self-energy})).

In general, the self-force acting on the particle at proper time
$\tau$ depends on the state of the particle at proper times both
earlier and later than $\tau$.
We can understand this by examining equation
(\ref{eqn:K-ret-extended-bar}) for $\bar{K}_r^\mu(\tau,\tau')$.
From the three factors in the integrand, it follows that
$\bar{K}_r^\mu(\tau,\tau')$ vanishes unless three conditions are met:
(1) there is an event $x^\mu$ within the world-cylinder of the
particle that lies on the plane of simultaneity for proper time
$\tau$,
(2) there is an event $x'^\mu$ within the world-cylinder of the
particle that lies on the plane of simultaneity for proper time
$\tau'$,
(3) events $x^\mu$ and $x'^\mu$ are light-like separated, with
$x'^\mu$ earlier than $x^\mu$.
There are retarded effects when $\bar{K}_r^\mu(\tau,\tau')$ is nonzero
for $\tau > \tau'$, and advanced effects when it is nonzero for
$\tau < \tau'$ (see Figure
\ref{fig:spacetime-diagrams-electrodynamics}).

We can give a concrete example that illustrates these conditions by
considering a uniformly accelerated particle in $(1+1)$ dimensions.
The position, velocity, and acceleration of the particle are given by
\begin{eqnarray}
  z^\mu(\tau) = a^{-1}\,e_1^\mu(\tau), \qquad
  v^\mu(\tau) = e_0^\mu(\tau), \qquad
  a^\mu(\tau) = a\,e_1^\mu(\tau),
\end{eqnarray}
where
\begin{eqnarray}
  e_0^\mu(\tau) \equiv (\cosh a \tau,\, \sinh a \tau), \qquad
  e_1^\mu(\tau) \equiv (\sinh a \tau,\, \cosh a \tau).
\end{eqnarray}
Note that $e_0 \cdot e_0 = - e_1 \cdot e_1 = 1$ and
$e_0 \cdot e_1 = 0$.
It is convenient to define a new coordinate system $(\lambda,u)$ by
\begin{eqnarray}
  x^\mu(\lambda, u) =
  z^\mu(\lambda) + u\,e_1^\mu(\lambda) =
  (a^{-1} + u)\,e_1^\mu(\lambda).
\end{eqnarray}
The trajectory of the particle is then given by
$z^\mu(\tau) = x^\mu(\tau,0)$, and the plane of simultaneity for
proper time $\tau$ is given by $x^\mu(\tau,u)$; note that the planes
of simultaneity for all values of $\tau$ pass through the origin at
$u = -1/a$.
Equation (\ref{eqn:j-bar}) for $\bar{J}^\mu(x,\tau)$ generalizes
naturally to $(1+1)$ dimensions; from the above expressions, it
follows that
\begin{eqnarray}
  \bar{J}^\mu(x,\tau) =
  e(1 + au)\,f(u^2)\,e_0^\mu(\tau)\,\delta(\lambda - \tau).
\end{eqnarray}
We will assume that $f(u^2) > 0$ for $u < R$ and $f(u^2) = 0$ for
$u > R$, so the trajectories of the left and right edges of the
particle are given by
$z_-^\mu(\lambda) \equiv x^\mu(\lambda, -R)$ and
$z_+^\mu(\lambda) \equiv x^\mu(\lambda, +R)$.

\begin{figure}[t]
  \centering
  \includegraphics{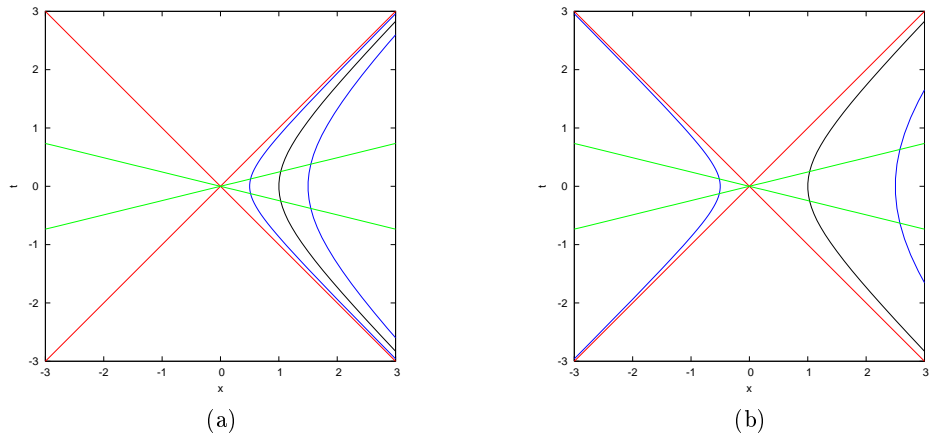}
  \caption{
    \label{fig:spacetime-diagram-uniform-acceleration}
    Spacetime diagrams for a uniformly accelerated particle of radius
    $R$.
    Black curve: particle trajectory $z^\mu$.
    Blue curves: trajectories $z_-^\mu$ and $z_+^\mu$ of
    the left and right edges of the particle.
    Red curves: trajectories $L_+^\mu$ and $L_-^\mu$.
    Green curves: planes of simultaneity for $\tau = \pm 1/4$.
    (a) $aR = 1/2$.
    (b) $aR = 3/2$.
  }
\end{figure}

If $aR < 1$, then the charge density is positive in the region between
the trajectories $z_-^\mu$ and $z_+^\mu$, and zero everywhere else
(see Figure \ref{fig:spacetime-diagram-uniform-acceleration}a).
Thus, conditions (1--3) are met only if $\tau > \tau'$.
If $aR > 1$, then the situation is more complicated.
Let us define trajectories
$L_\pm^\mu(\lambda) = (\lambda,\, \pm|\lambda|)$;
these trajectories define the left and right edges of the forward and
backward light-cones for the origin.
Also, let us define $R_+$ to be the region between $L_+^\mu$ and
$z_+^\mu$, and $R_-$ to be the region between $z_-^\mu$ and
$L_-^\mu$.
The charge density is positive in the region $R_+$, negative in the
region $R_-$, and zero everywhere else
(see Figure \ref{fig:spacetime-diagram-uniform-acceleration}b).
Thus, conditions (1--3) are met by events in region $R_+$ if
$\tau > \tau'$, and by events in region $R_-$ if $\tau < \tau'$.

\section{Toy model of electrodynamics}
\label{sec:toy-model}

We can gain further insight into the advanced effects by comparing
ordinary electrodynamics with a toy model of electrodynamics in which
advanced effects do not occur.
A complete description of the model is given in \cite{boozer-a}, but
all the results that we will need are summarized here.

The toy model that we will be considering describes a spatially
extended particle in $(1+1)$ dimensions that obeys Newtonian dynamics
and is coupled to pair of fields $E(t,x)$ and $B(t,x)$, which
correspond to the electric and magnetic fields of ordinary
electrodynamics.
The equations of motion for these fields are
\begin{eqnarray}
  \label{eqn:eqn-of-motion-E}
  \partial_t E(t,x) & = & \partial_x B(t,x), \\
  \label{eqn:eqn-of-motion-B}
  \partial_t B(t,x) & = & \partial_x E(t,x) - 2\rho(t,x),
\end{eqnarray}
where $\rho(t,x)$ is the charge density.
We will assume that the charge density has the form
\begin{eqnarray}
  \label{eqn:charge-density}
  \rho(t,x) = g\,f(x - z(t)),
\end{eqnarray}
where $z(t)$ is the position of the particle at time $t$, $g$ is its
charge, and $f(x)$ describes the charge distribution.
The equation of motion for the particle is
\begin{eqnarray}
  \label{eqn:eqn-of-motion-particle}
  m\ddot{z} = F_f + F_{ext},
\end{eqnarray}
where $m$ is the particle mass,
\begin{eqnarray}
  \label{eqn:lorentz-force}
  F_f(t) = -2\int \rho(t,x)E(t,x)\,dx
\end{eqnarray}
is the force that the $E$-field exerts on the particle, and
$F_{ext}$ describes an arbitrary externally imposed force.
Equations (\ref{eqn:eqn-of-motion-E}--\ref{eqn:lorentz-force}) give a
complete description of the coupled particle-field system.
Equations (\ref{eqn:eqn-of-motion-E}) and (\ref{eqn:eqn-of-motion-B})
can be though of as the analogs to Maxwell's equations, and equation
(\ref{eqn:lorentz-force}) can be thought of as the analog to the
Lorentz force law.

By analogy with electrodynamics, we can express the $E$ and $B$
fields in the form
\begin{eqnarray}
  \label{eqn:decompositions}
  E(t,x) = E_r(t,x) + E_{in}(t,x), \qquad
  B(t,x) = B_r(t,x) + B_{in}(t,x),
\end{eqnarray}
where $E_r(t,x)$ and $B_r(t,x)$ are the retarded fields generated
by the particle, and $E_{in}(t,x)$ and $B_{in}(t,x)$ describe
incoming radiation (see \cite{boozer-b}).
The retarded fields are given by
\begin{eqnarray}
  \label{eqn:E-r-B-r}
  E_r(t,x) = \partial_x\phi_r(t,x), \qquad
  B_r(t,x) = \partial_t\phi_r(t,x),
\end{eqnarray}
where
\begin{eqnarray}
  \label{eqn:phi-r}
  \phi_r(t,x) =
  \int\!\!\!\int D_r( t-t', x-x')\,\rho(t',x')\,dt'\,dx'
\end{eqnarray}
and $D_r(t,x) = \theta(t - |x|)$
is the retarded Green function for the inhomogeneous wave equation in
$(1+1)$ dimensions.
Using the decompositions given in equation (\ref{eqn:decompositions}),
we can express the force exerted by the field as
$F_f = F_{in} + F_r$, where
\begin{eqnarray}
  \label{eqn:F-r-F-in}
  F_{in}(t) = -2\int \rho(t,x) E_{in}(t,x)\,dx, \qquad
  F_r(t) = -2\int \rho(t,x) E_r(t,x)\,dx.
\end{eqnarray}
Physically, $F_{in}$ describes the force exerted on the particle by
incoming radiation, and $F_r$ describes the self-force exerted on the
particle by its own retarded field.
In what follows we will assume that there is no incoming radiation, so
$F_{in} = 0$ and $F_f = F_r$.

Using equations (\ref{eqn:E-r-B-r}), (\ref{eqn:phi-r}), and
(\ref{eqn:F-r-F-in}), we can evaluate the self-force explicitly for
the case of a point particle, for which $f(x) = \delta(x)$:
\begin{eqnarray}
  \label{eqn:F-reaction-point}
  F_r = -m\gamma \dot{z} (1 - \dot{z}^2)^{-1},
\end{eqnarray}
where $\gamma \equiv 2g^2/m$ is a damping constant.
From equations (\ref{eqn:eqn-of-motion-particle}) and
(\ref{eqn:F-reaction-point}), we find that the equation of
motion is
\begin{eqnarray}
  \label{eqn:lorentz-dirac-toymodel}
  \ddot{z} + \gamma \dot{z} (1 - \dot{z}^2)^{-1} = (1/m)F_{ext}.
\end{eqnarray}
This is the toy model analog to the Lorentz-Dirac equation
(\ref{eqn:lorentz-dirac}) for electrodynamics.
If we compare equation (\ref{eqn:lorentz-dirac-toymodel}) with the
Lorentz-Dirac equation, we note three important differences.
First, the Lorentz-Dirac equation is third order, but
equation (\ref{eqn:lorentz-dirac-toymodel}) is only second order, and
thus does not admit runaway solutions and does not need to be replaced
with an integro-differential equation.
In the toy model the acceleration of a point particle at time $t$ only
depends on the value of $F_{ext}$ at time $t$, and there are no
advanced effects.
Second, for the Lorentz-Dirac equation there is a radiation damping
term proportional to the time-derivative of the acceleration, while
for equation (\ref{eqn:lorentz-dirac-toymodel}) the damping term is
a function of the particle velocity.
For electrodynamics, a velocity-dependent damping term is
ruled out by Lorentz invariance, but the toy model is neither Lorenz
nor Galilean invariant; there is a preferred reference frame in which
the equations of motion for the model are valid, and a particle moving
with respect to this preferred frame feels a velocity-dependent drag
force.
Third, the mass that appears in the Lorentz-Dirac equation is the
renormalized mass, the sum of the bare mass and the self-energy, while
the mass that appears in equation (\ref{eqn:lorentz-dirac-toymodel})
is just the bare mass; there is no self-energy contribution.
We will explain the reason for this in section
\ref{ssec:eqn-of-motion-approximate}.

\section{Extended particles in the toy model}
\label{sec:toy-model-advanced}

\begin{figure}
  \centering
  \includegraphics{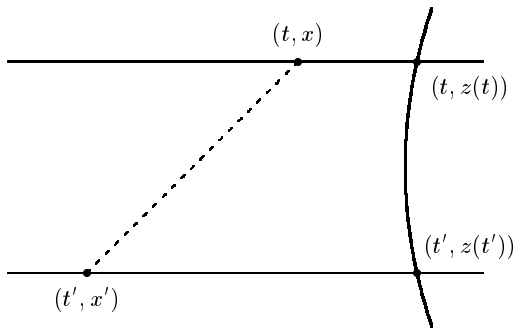}
  \caption{
    \label{fig:spacetime-diagram-toymodel}
    Spacetime diagram, which illustrates a pair of events $(t,x)$ and
    $(t',x')$ that give a nonzero contribution to $\bar{F}_r(t,t')$.
    The thick solid line indicates the particle trajectory,
    the thin solid lines indicate the planes of simultaneity for times
    $t$ and $t'$,
    the dashed line indicates the light-like interval between events
    $(t',x')$ and $(t,x)$.
  }
\end{figure}

Let us now consider a spatially extended particle in the toy model.
From equations (\ref{eqn:E-r-B-r}), (\ref{eqn:phi-r}), and
(\ref{eqn:F-r-F-in}), it follows that the self-force for an extended
particle is given by
\begin{eqnarray}
  \label{eqn:F-ret-extended}
  F_r(t) = \int \bar{F}_r(t,t')\,dt',
\end{eqnarray}
where we have defined
\begin{eqnarray}
  \label{eqn:F-ret-extended-bar}
  \bar{F}_r(t,t') =
  -2 \int\!\!\!\int
  \rho(t,x)\,\partial_x D_r(t-t', x-x')\,\rho(t',x')\,dx\,dx'.
\end{eqnarray}
Equations
(\ref{eqn:F-ret-extended}) and
(\ref{eqn:F-ret-extended-bar}) are the toy model analogs of equations
(\ref{eqn:K-ret-extended}) and
(\ref{eqn:K-ret-extended-bar}) for ordinary electrodynamics.
Note, however, that whereas $\bar{K}_r(\tau, \tau')$ can be nonzero
for either $\tau > \tau'$ or $\tau < \tau'$, the analogous quantity
$\bar{F}_r(t,t')$ is only nonzero if $t > t'$.
Physically, this is due to the fact that the particle obeys Newtonian
dynamics, so the planes of simultaneity do not tilt (see Figure
\ref{fig:spacetime-diagram-toymodel}, and compare with
Figure \ref{fig:spacetime-diagrams-electrodynamics} for
electrodynamics).
In what follows, we will assume that the charge distribution of the
extended particle is
\begin{eqnarray}
  \label{eqn:gaussian-profile}
  f(x) = (2\pi\sigma^2)^{-1/2}\,e^{-x^2/2\sigma^2},
\end{eqnarray}
where $\sigma$ describes the particle size.
If we substitute equation (\ref{eqn:gaussian-profile}) into equation
(\ref{eqn:F-ret-extended-bar}), we find that
\begin{eqnarray}
  \bar{F}_r(t,t') =
  (m\gamma/2\sqrt{\pi}\sigma)\,\theta(t - t')\,
  (e^{-(z(t) - z(t') + t - t'))^2/4\sigma^2} -
  e^{-(z(t) - z(t') + t' - t)^2/4\sigma^2}).
\end{eqnarray}

The equation of motion for an extended particle is
\begin{eqnarray}
  \label{eqn:lorentz-dirac-extended-toymodel}
  m\ddot{z} = F_r + F_{ext},
\end{eqnarray}
where $F_r$ is given by equation (\ref{eqn:F-ret-extended}) and
$F_{ext}$ describes an externally imposed force.
This is the toy model analog to equation
(\ref{eqn:lorenz-dirac-extended}), the equation of motion for an
extended particle in electrodynamics.
Like equation (\ref{eqn:lorenz-dirac-extended}), it is a
second-order integro-differential equation that is exact, and holds
for extended particles of arbitrary size.
Given the velocity $v_0$ at time $t_0$, together with the particle
trajectory $z(t)$ at all times $t \leq t_0$, equation
(\ref{eqn:lorentz-dirac-extended-toymodel}) can be integrated to
obtain the particle trajectory at all times.

\subsection{Approximate equation of motion}
\label{ssec:eqn-of-motion-approximate}

We can further develop the analogy between the toy model and
electrodynamics by performing a series expansion of the self-force for
an extended particle.
If we substitute equation
(\ref{eqn:F-ret-extended-bar}) into equation
(\ref{eqn:F-ret-extended}) and perform the integral over $t'$, we find
that
\begin{eqnarray}
  \label{eqn:F-reaction-def}
  F_r(t) =
  -2\int\!\!\!\int \epsilon(x - x')\,\rho(t,x)\,
  \rho(t - |x - x'|,x')\,dx\,dx'.
\end{eqnarray}
Let us expand $\rho(t - |x - x'|, x')$ in $|x- x'|$:
\begin{eqnarray}
  \label{eqn:series-a}
  \rho(t - |x - x'|, x') =
  \sum_{n=0}^\infty (1/n!)(-1)^n\,
  |x - x'|^n\,\partial_t^{n}\rho(t,x').
\end{eqnarray}
If we assume that the particle is moving slowly, then we can neglect
terms that are nonlinear in $z^{(n)}(t) \equiv d^n z(t)/dt^n$:
\begin{eqnarray}
  \label{eqn:series-b}
  \rho(t - |x - x'|, x') =
  \rho(t,x') -
  \sum_{n=1}^\infty (1/n!)(-1)^n\,|x - x'|^n\,z^{(n)}(t)\,
  \partial_{x'}\rho(t,x').
\end{eqnarray}
Substituting equation (\ref{eqn:series-b}) into equation
(\ref{eqn:F-reaction-def}), we find that
\begin{eqnarray}
  \label{eqn:F-reaction-series}
  F_r(t) = -m\gamma\sum_{n=0}^\infty (-1)^n\,c_n \sigma^n v^{(n)}(t),
\end{eqnarray}
where $v^{(n)} \equiv d^n v(t)/dt^n$, $v = \dot{z}$ is the particle
velocity, and
\begin{eqnarray}
  \label{eqn:cn-general}
  c_n \equiv
  (1/n!\,\sigma^n) \int\!\!\!\int |x - x'|^n f(x) f(x')\,dx\,dx' =
  \frac{2^n}{\sqrt{\pi}}\frac{\Gamma(n/2 + 1/2)}{\Gamma(n + 1)}
\end{eqnarray}
are dimensionless coefficients.
The first few terms of equation (\ref{eqn:F-reaction-series}) are:
\begin{eqnarray}
  \label{eqn:F-reaction-first-few}
  F_r = -m\gamma v - m_S \dot{v} + m\tau_0\ddot{v} + \cdots,
\end{eqnarray}
where $m_S \equiv -(4/\sqrt{\pi})g^2\sigma$ and
$\tau_0 \equiv -2 g^2 \sigma^2/m$.
The first term is just the low-velocity limit of the self-force
for a point particle, and the second term describes the self-energy
$m_S$.
Since the self-energy is proportional to the particle size, it
vanishes in the point particle limit, which explains why it is the
bare mass, rather than the renormalized mass, that appears in the
point particle equation of motion (\ref{eqn:lorentz-dirac-toymodel}).

It is instructive to compare equation (\ref{eqn:F-reaction-first-few})
with the the low-velocity limit of the electrodynamic self-force
given in equation (\ref{eqn:K-r}):
\begin{eqnarray}
  \label{eqn:K-r-nonrel}
  \mathbf{K}_r = -m_S \mathbf{a} + m \tau_0 \dot{\mathbf{a}} + \cdots,
\end{eqnarray}
where the dots indicate terms that vanish in the point particle limit.
Note that for electrodynamics the term proportional to the velocity
is not present (as we discussed before, such a term is ruled out by
Lorentz invariance).
Also, the self-energy $m_S$ and time constant $\tau_0$ are both
negative for the toy model, but positive for electrodynamics.
These quantities also scale differently in the two theories:
for the toy model, $m_S \sim \sigma$ and $\tau_0 \sim \sigma^2$,
whereas for electrodynamics $m_S \sim 1/\sigma$ and $\tau_0$ is
independent of the particle size.
Thus, in the point particle limit,
\begin{eqnarray}
  \label{eqn:F-r-K-r-point-particle}
  F_r \rightarrow -m\gamma v, \qquad
 \mathbf{K}_r \rightarrow -m_S \mathbf{a} + m \tau_0 \dot{\mathbf{a}}.
\end{eqnarray}

For a small enough particle, we can obtain a good approximation to
equation (\ref{eqn:F-reaction-series}) by truncating the series
at some finite order $N$:
\begin{eqnarray}
  \label{eqn:reaction-force-approximate}
  F_r = -m\gamma\sum_{n=0}^{N-1} (-1)^n\,c_n \sigma^n v^{(n)}.
\end{eqnarray}
Thus, we obtain an approximate equation of motion for the particle:
\begin{eqnarray}
  \label{eqn:eqn-of-motion-approximate}
  v^{(1)} + \gamma\sum_{n=0}^{N-1} (-1)^n\,c_n \sigma^n v^{(n)} =
  (1/m)F_{ext}.
\end{eqnarray}
This should be a good approximation to the exact equation of motion
(\ref{eqn:lorentz-dirac-extended-toymodel}) provided the particle is
small and slowly-moving.
Since this equation is of order $N$, we need initial
conditions $z^{(n)}(0)$ for $n=0,1,\cdots, N-1$.

We can write down the solutions to equation
(\ref{eqn:eqn-of-motion-approximate}) for the special case
$F_{ext} = 0$:
\begin{eqnarray}
  v(t) = \sum_{k=1}^{N-1} A_k\,e^{-\beta_k t/\sigma},
\end{eqnarray}
where the constants $A_1, \cdots, A_{N-1}$ are set by the initial
conditions, the constants $\beta_1, \cdots, \beta_{N-1}$ are the $N-1$
roots of the polynomial
\begin{eqnarray}
  \label{eqn:beta}
  p(\beta) = \beta  - \eta \sum_{n=0}^{N-1} c_n \beta^n,
\end{eqnarray}
and $\eta \equiv \gamma\sigma$ is a dimensionless measure of the
particle size.
Note that some of the roots may be complex, and if so then the initial
conditions must be chosen such that $v(t)$ is real
(the fact that there are complex roots suggests that there are
oscillatory solutions to the equation of motion; examples of such
solutions are given in Appendix \ref{sec:appendix-oscillations}).
Also, note that if $\textup{Re}\,\beta_k < 0$ and $A_k \neq 0$, then
we obtain a runaway solution in which the velocity increases
exponentially in time.
We can write down the roots explicitly in the limit of small particle
size ($\eta \ll 1$):
\begin{eqnarray}
  \label{eqn:beta-k}
  \beta_{N-1} = \eta, \qquad \qquad
  \beta_k = e^{2\pi i(k-1)/(N-2)}\,(\eta c_{N-1})^{-1/(N-2)}
  \quad \textup{for $k=1,\cdots,N-2$}.
\end{eqnarray}
Note that for $N \geq 4$ there is at least one root that yields a
runaway solution.
For the case $N=3$ we can calculate the roots exactly and write down
the solutions explicitly; this is done in Appendix
\ref{sec:appendix-solutions}.

\subsection{Integro-differential approximate equation of motion}
\label{ssec:int-diff-eqn}

As for electrodynamics, we can eliminate the runaway solutions by
replacing the approximate equation of motion
(\ref{eqn:eqn-of-motion-approximate}) with an integro-differential
equation.
Let us define a Green function $G(t)$ by
\begin{eqnarray}
  \label{eqn:greens-function-def}
  G^{(1)}(t) + \gamma\sum_{n=0}^{N-1}
  (-1)^n\,c_n\,\sigma^n\,G^{(n)}(t) = \delta(t).
\end{eqnarray}
We will assume that $N \geq 3$, so that the polynomial $p(\beta)$
defined in equation (\ref{eqn:beta}) can be factorized as follows:
\begin{eqnarray}
  p(\beta) =
  \beta - \eta\sum_{n=0}^{N-1} c_n \beta^n =
  -\eta c_{N-1}\prod_{n=1}^{N-1} (\beta - \beta_n).
\end{eqnarray}
Using this result, it is straightforward to show that
the Fourier transform $\tilde{G}(\omega)$ of $G(t)$ is
\begin{eqnarray}
  \tilde{G}(\omega) =
  \int G(t)\,e^{i\omega t}\,dt =
  (c_{N-1} \gamma)^{-1}\prod_{k=1}^{N-1} (i\omega\sigma - \beta_k)^{-1}.
\end{eqnarray}
Thus, $G(t)$ is given by
\begin{eqnarray}
  \label{eqn:greens-function-eval}
  G(t) =
  (2\pi)^{-1}\int \tilde{G}(\omega)\,e^{-i\omega t}\,d\omega =
  \sum_{k=1}^{N-1} B_k\,e^{-\beta_k t/\sigma}\,
  \theta(\epsilon_k t),
\end{eqnarray}
where we have defined constants $B_1, \cdots, B_{N-1}$ and
$\epsilon_1, \cdots, \epsilon_{N-1}$ by
\begin{eqnarray}
  B_k \equiv -(\epsilon_k/\eta c_{N-1})
  \prod_{j \neq k} (\beta_k - \beta_j)^{-1}, \qquad
  \epsilon_k \equiv
  \left\{
  \begin{array}{ll}
    +1 & \quad \mbox{if $\textup{Re}\,\beta_k > 0$} \\
    -1 & \quad \mbox{if $\textup{Re}\,\beta_k < 0$}
  \end{array}
  \right.
\end{eqnarray}

Using the Green function, we can express the solution to the
approximate equation of motion (\ref{eqn:eqn-of-motion-approximate})
as an integro-differential equation:
\begin{eqnarray}
  \label{eqn:approximate-solution}
  v(t) = (1/m)\int G(t-t')\,F_{ext}(t')\,dt' =
  (1/m)\sum_{k=1}^{N-1} B_k\,
  \int \theta(\epsilon_k \tau)\,e^{-\beta_k \tau/\sigma}\,
  F_{ext}(t - \tau)\,d\tau.
\end{eqnarray}
As for electrodynamics, by expressing the solution
to a higher-order equation of motion in terms of a suitable Green
function we have eliminated the runaway solutions; for example,
equation (\ref{eqn:approximate-solution}) implies that
$v(t) = 0$ for $F_{ext} = 0$.
Note that terms with $\epsilon_k = 1$ only depend on the value of the
external force at times $t' < t$, and thus describe retarded effects,
while terms with $\epsilon_k = -1$
only depend on the value of the external force at times $t' > t$, and
thus describe advanced effects.
The advanced effects are not present for the exact equation of
motion (\ref{eqn:lorentz-dirac-extended-toymodel}); rather, they are
an artifact of the approximations used to obtain equation
(\ref{eqn:approximate-solution}).

From equation (\ref{eqn:beta-k}), it follows that in the limit of
small particle size there will be advanced effects if $N \geq 4$.
Let us compare this with electrodynamics.
Recall that the low-velocity limit of the electrodynamic self-force is
given by equation (\ref{eqn:K-r-nonrel}).
It is the point particle limit of this self-force that appears in the
equation of motion (\ref{eqn:abraham-lorentz}), which, as we showed in
section \ref{sec:lorentz-dirac}, gives rise to advanced effects.
But taking the point particle limit of equation (\ref{eqn:K-r-nonrel})
is equivalent to truncating a series expansion of the self-force
at order $N=3$.
Thus, advanced effects show up at third order for electrodynamics, but
only at fourth order for the toy model.
This can be understood by comparing equations
(\ref{eqn:F-reaction-first-few}) and (\ref{eqn:K-r-nonrel}) for the
self-forces in the two theories.
For both theories, the sign of the third-order term is determined by
the sign of $\tau_0$: for electrodynamics $\tau_0$ is positive,
corresponding to advanced effects, while for the toy model $\tau_0$ is
negative, corresponding to retarded effects.
Also, note that for the toy model the higher-order terms in the
self-force vanish in the point particle limit, and thus the spurious
advanced effects exhibited by equation
(\ref{eqn:approximate-solution}) go away in this limit.
For electrodynamics, however, the third-order term of the self-force
is independent of the particle size, and hence the advanced effects
remain in the point particle limit.

\begin{figure}
  \centering
  \includegraphics{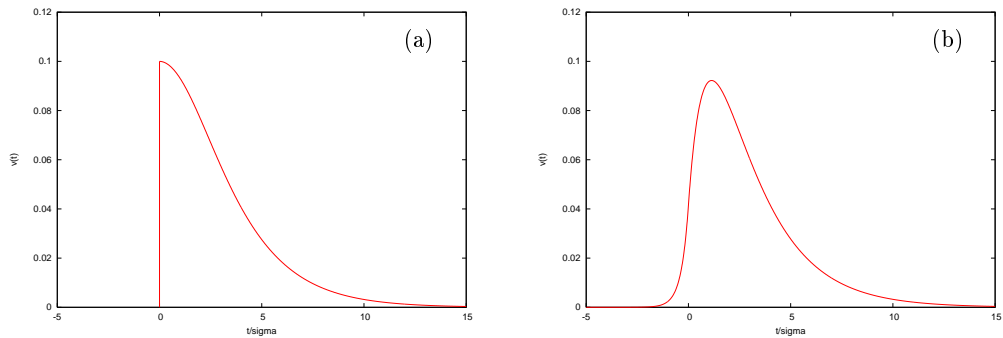}
  \caption{
    \label{fig:solution}
    Graph of $v(t)$ versus $t/\sigma$ for an impulsive force
    ($\eta = 0.25$, $v_0 = 0.1$).
    (a) Exact equation of motion
    (\ref{eqn:lorentz-dirac-extended-toymodel}).
    (b) Approximate integro-differential equation of motion
    (\ref{eqn:approximate-solution}) with $N=4$.
  }
\end{figure}

As an example, let us consider the special case of an impulsive force
$F_{ext}(t) = m v_0 \delta(t)$, and compare the evolution predicted by
the exact equation of motion
(\ref{eqn:lorentz-dirac-extended-toymodel}) with the evolution
predicted by the approximate integro-differential equation of motion
(\ref{eqn:approximate-solution}).
Suppose that the particle starts at rest at the origin, so $z(t) = 0$
for $t \leq 0$.
The moment after the impulsive force is applied, the velocity of the
particle is $v(0) = v_0$.
We can numerically integrate the exact equation of motion
(\ref{eqn:lorentz-dirac-extended-toymodel}) for these initial
conditions; the result is shown in Figure \ref{fig:solution}a, where
we have taken $\eta = 0.25$ and $v_0 = 0.1$.
As expected, there are no advanced effects: the particle does not
respond to the impulsive force until after it has been applied.
We can also describe the evolution of the particle using the
approximate integro-differential equation
(\ref{eqn:approximate-solution}).
If we substitute for $F_{ext}(t)$, we find that
\begin{eqnarray}
  v(t) = v_0\sum_{k=1}^{N-1}
  B_k\,e^{-\beta_k t/\sigma}\,\theta(\epsilon_k t).
\end{eqnarray}
In Figure \ref{fig:solution}b this solution is shown for
$\eta = 0.25$, $v_0 = 0.1$, $N=4$.
Now there are advanced effects: the particle begins to accelerate
before the impulsive force has been applied.

\section{Conclusion}

We have shown that the reason for the advanced effects in classical
electrodynamics can be traced to the fact that the planes of
simultaneity for an extended particle can tilt in such a way that the
self-force exerted on the particle at proper time $\tau$ depends on
the state of the particle at proper times $\tau' > \tau$.
We have also considered a toy model of electrodynamics in which the
planes of simultaneity do not tilt, and have shown that it does not
give rise to advanced effects.

We have written down three equations of motion for electrodynamics.
Equation (\ref{eqn:lorenz-dirac-extended}) is an exact equation of
motion for an extended particle; it is a second-order
integro-differential equation that exhibits advanced effects.
Equation (\ref{eqn:lorentz-dirac}), the Lorentz-Dirac equation,
approximates equation (\ref{eqn:lorenz-dirac-extended}) in the
limit of a small particle; it is a third order ordinary differential
equation, and admits runaway solutions.
Equation (\ref{eqn:lorentz-dirac-int-diff}) is obtained
from the Lorentz-Dirac equation by choosing the initial conditions so
as to eliminate the runaway solutions.
It also approximates equation (\ref{eqn:lorenz-dirac-extended}) in the
limit of a small particle, and like equation
(\ref{eqn:lorenz-dirac-extended}) it is a second-order
integro-differential equation that exhibits advanced effects.

We have written down four equations of motion for the toy model, which
can be viewed as analogs to the various electrodynamic equations of
motion.
Equation (\ref{eqn:lorentz-dirac-extended-toymodel}) is an exact
equation of motion for an extended particle; it is a second-order
integro-differential equation, and does not exhibit advanced effects.
Equation (\ref{eqn:lorentz-dirac-toymodel}) is an exact equation of
motion for a point particle; it is a second-order ordinary
differential equation, and does not exhibit advanced effects.
Equation (\ref{eqn:eqn-of-motion-approximate}) approximates equation
(\ref{eqn:lorentz-dirac-extended-toymodel}) in the limit of a small
particle; it is an ordinary differential equation of order $N$, and
admits runaway solutions for $N \geq 4$.
Equation (\ref{eqn:approximate-solution}) is obtained from equation
(\ref{eqn:eqn-of-motion-approximate}) by choosing the initial
conditions so as to eliminate the runaway solutions; it is a
first-order integro-differential equation, and exhibits spurious
advanced effects for $N \geq 4$.

\appendix

\section{Self-energy of an extended particle}
\label{sec:appendix-self-energy}

In this Appendix we show that the self-force given in equation
(\ref{eqn:K-ret-extended}) can be used to obtain the correct
expression for the self-energy of a spatially extended particle.
To accomplish this, we will use equation (\ref{eqn:K-ret-extended}) to
calculate the self-force that acts on a spatially extended particle
undergoing uniformly accelerated motion.
For simplicity we will work to first order in the acceleration, and
calculate the self-force at the moment at which the particle is
instantaneously at rest.
The trajectory, velocity, and acceleration of the particle are
\begin{eqnarray}
  z^\mu(\tau) = (\tau,\, \mathbf{a}\tau^2/2), \qquad
  v^\mu(\tau) = (1,\, \mathbf{a}\tau), \qquad
  a^\mu(\tau) = (0,\, \mathbf{a}).
\end{eqnarray}
From equation (\ref{eqn:j-bar}), we find that
\begin{eqnarray}
  \bar{J}^\mu(x, \tau) =
  e (1 + \mathbf{a}\cdot\mathbf{x})
  f(|\mathbf{x} - \mathbf{z}(\tau)|^2)\,v^\mu(\tau)\,
  \delta(t - (1 + \mathbf{a}\cdot\mathbf{x})\tau),
\end{eqnarray}
where $t \equiv x^0$.
The current density is given by
\begin{eqnarray}
  J^\mu(x) =
  \int \bar{J}^\mu(x, \tau)\,d\tau =
  e \int
  f(|\mathbf{x} - \mathbf{z}(\tau)|^2)\,v^\mu(\tau)\,
  \delta(\tau - (1 + \mathbf{a}\cdot\mathbf{x})^{-1}\,t)\,d\tau,
\end{eqnarray}
so
\begin{eqnarray}
  J^0(x) = \rho(t,\mathbf{x}), \qquad
  \mathbf{J}(x) = \mathbf{a}t\,\rho(t,\mathbf{x}),
\end{eqnarray}
where we have defined
\begin{eqnarray}
  \rho(t,\mathbf{x}) \equiv e\,f(|\mathbf{x} - \mathbf{a}t^2/2|^2).
\end{eqnarray}
Also, note that
\begin{eqnarray}
  \label{eqn:j-bar-0}
  \bar{J}^\mu(x,0) =
  (1 + \mathbf{a}\cdot\mathbf{x})\,\rho(0,\mathbf{x})\,
  {\delta^\mu}_0\,\delta(t - \tau).
\end{eqnarray}
Substituting equation (\ref{eqn:j-bar-0}) into equation
(\ref{eqn:K-ret-extended}), we find that the self-force at $\tau=0$ is
\begin{eqnarray}
  \label{eqn:self-force-uniform-acceleration}
  \mathbf{K}_r(0) =
  \int \rho(0,\mathbf{x})\,\mathbf{E}_r(0,\mathbf{x})\,d^3 x +
  \int \mathbf{a}\cdot\mathbf{x}\,
  \rho(0,\mathbf{x})\,\mathbf{E}_r(0,\mathbf{x})\,d^3 x,
\end{eqnarray}
where $\mathbf{E}_r(t,\mathbf{x})$, the retarded electric field, is
given by
\begin{eqnarray}
  \mathbf{E}_r(t,\mathbf{x}) =
  -\nabla A^0_r(t,\mathbf{x}) - \partial_t \mathbf{A}_r(t,\mathbf{x}),
\end{eqnarray}
and $A_r^\mu(x)$, the retarded vector potential, can be obtained from
the current density $J^\mu(x)$ via equation (\ref{eqn:A-ret-adv}).
After a lengthy but straightforward calculation, we find that
\begin{eqnarray}
  \mathbf{E}_r(t,\mathbf{x}) =
  -\nabla \int
  |\mathbf{x} - \mathbf{x}'|^{-1}\,\rho(t,\mathbf{x}')\,d^3 x' -
  (2/3)\,\mathbf{a}\int
  |\mathbf{x} - \mathbf{x}'|^{-1}\,\rho(t,\mathbf{x}')\,d^3 x'.
\end{eqnarray}
Substituting this result into equation
(\ref{eqn:self-force-uniform-acceleration}), we find that the
self-force is
\begin{eqnarray}
  \mathbf{K}_r(0) = -m_S \mathbf{a},
\end{eqnarray}
where the self-energy $m_S$ is given by
\begin{eqnarray}
  \label{eqn:self-energy}
  m_S =
  (e^2/2)\int\!\!\!\int |\mathbf{x} - \mathbf{x}'|^{-1}
  f(|\mathbf{x}|^2)\,f(|\mathbf{x}'|^2)\,d^3x\,d^3x'.
\end{eqnarray}
It is interesting to note that the Abraham-Lorentz self-force
corresponds to just the first term of equation
(\ref{eqn:self-force-uniform-acceleration}), and gives an
incorrect value of $(4/3)m_S$ for the self-energy
(see \cite{lorentz}, and section 17.3 of \cite{jackson}).
The correct relativistic expression for the self-force given in
equation (\ref{eqn:K-ret-extended}) modifies this result by
producing the second term of equation
(\ref{eqn:self-force-uniform-acceleration}), which evaluates to
$-(1/3)m_S$ and combines with the first term to give the correct
value for the self-energy.

\section{Example solutions for a spatially extended particle}
\label{sec:appendix-oscillations}

\begin{figure}[t]
  \centering
  \includegraphics[scale=0.25,angle=270]{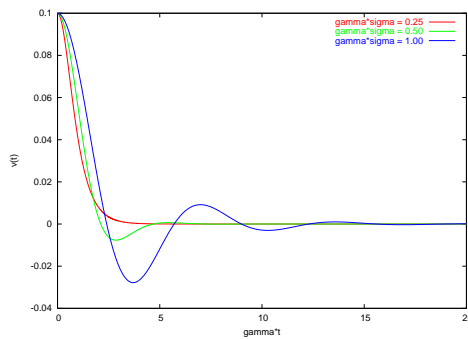}
  \caption{
    \label{fig:oscillatory-solution}
    Graph of $v(t)$ versus $\gamma t$ for an impulsive force
    ($v_0=0.1$).
    Red curve: $\gamma\sigma = 0.25$.
    Green curve: $\gamma\sigma = 0.5$.
    Blue curve: $\gamma\sigma = 1.0$.
  }
\end{figure}

In this Appendix we consider some example solutions to equation
(\ref{eqn:lorentz-dirac-extended-toymodel}), the exact equation of
motion for an extended particle in the toy model.
Note that one region of an extended particle can cause a
change in the field that acts back on a different region of the
particle at a later time; as we shall see, if the particle is large
enough the delay between these events can lead to oscillatory
behavior.

As for the example in section \ref{ssec:int-diff-eqn}, we will assume
that the  particle is initially at rest at the origin, and that it is
driven with an impulsive force $F_{ext}(t) = m v_0 \delta(t)$; this is
equivalent to taking the initial conditions of the particle to be
$v(0) = v_0$, $z(t) = 0$ for $t \leq 0$.
We numerically integrate the equation of motion
(\ref{eqn:lorentz-dirac-extended-toymodel}) subject to these initial
conditions, and plot $v(t)$ versus $\gamma t$ in Figure
\ref{fig:oscillatory-solution}.
Curves are shown for three different values of the particle size
$\sigma$.
Note that the renormalized mass of an extended particle is
$m_R =  m + m_S = (1 - 2\gamma\sigma/\sqrt{\pi})m$, so for the curve
with $\gamma\sigma = 1$ the renormalized mass is negative.

\section{Solution for $N=3$}
\label{sec:appendix-solutions}

Here we find the solutions to equation
(\ref{eqn:eqn-of-motion-approximate}), the approximate equation of
motion for the toy model, for the case $N=3$.
We will assume there is no external driving force ($F_{ext} = 0$), so
we can express the equation of motion as
\begin{eqnarray}
  \dot{v} = -\gamma_R v + \tau_R \ddot{v},
\end{eqnarray}
where $\tau_R \equiv (m/m_R)\tau_0$, $\gamma_R \equiv (m/m_R)\gamma$,
$m_R \equiv m + m_S$.
The solutions are given by
\begin{eqnarray}
  v(t) = (\alpha_+ - \alpha_-)^{-1}
  ((\alpha_+ v_0 + a_0)\,e^{-\alpha_- t} -
  (\alpha_- v_0 + a_0)\,e^{-\alpha_+ t}),
\end{eqnarray}
where
$\alpha_\pm \equiv -(1/2\tau_R)(1 \pm (1 + 4\gamma_R\tau_R)^{1/2})$,
and $v_0$ and $a_0$ are the initial velocity and acceleration.

\end{document}